# Suppression of the antiferromagnetic metallic state in the pressurized MnBi$_2$Te$_4$ single crystal

K. Y. Chen,[1,5] B. S. Wang,[1,5,6] J.-Q. Yan,[2] D. S. Parker,[2] J.-S. Zhou,[3] Y. Uwatoko,[4] and J.-G. Cheng[1,4,5,6]∗

[1]*Beijing National Laboratory for Condensed Matter Physics and Institute of Physics, Chinese Academy of Sciences, Beijing, China 100190*

[2]*Materials Science and Technology Division, Oak Ridge National Laboratory, Oak Ridge, Tennessee 37831, USA*

[3]*Materials Science and Engineering Program, Mechanical Engineering, University of Texas at Austin, Austin, Texas 78712, USA*

[4]*Institute for Solid State Physics, University of Tokyo, Kashiwanoha 5-1-5, Kashiwa, Chiba 277-8581, Japan*

[5]*School of Physical Sciences, University of Chinese Academy of Sciences, Beijing, China 100190*

[6]*Songshan Lake Materials Laboratory, Dongguan, Guangdong, China 523808*

E-mail: jgcheng@iphy.ac.cn

## Abstract

MnBi$_2$Te$_4$ has attracted tremendous research interest recently as the first intrinsic antiferromagnetic (AF) topological insulator. It undergoes a long-range AF order at $T_N$ ≈ 24 K accompanied with a cusp-like anomaly in the metallic resistivity. Here, we studied the effect of hydrostatic pressure on its electrical transport properties up to 12.5 GPa by using a cubic anvil cell apparatus. We find that $T_N$ determined from the resistivity anomaly first increases slightly with pressure and then decreases until vanished completely at ~7 GPa. Intriguingly, its resistivity $\rho(T)$ is enhanced gradually by pressure, and evolves from metallic to activated behavior as the AF order is suppressed. From the Hall resistivity measurements, we confirm that the n-type carriers dominate the transport properties and the carrier density is raised by pressure. In addition, the critical magnetic field $H_{c1}$ ~3.3 T at 0 GPa for the spin-flop transition to the canted AF state is found to increase to ~ 5 T and 7.5 T at 1 and 3 GPa. High-pressure XRD evidenced no structural transition up to 12.8 GPa. Based on the Hall resistivity results and first-principles calculations, we proposed that the intralayer direct AF interactions are strengthened by pressure and the competition between AF and FM interactions not only prevents long-range magnetic order but also promotes charge carrier localizations through enhance magnetic fluctuations at high pressures.

## Introduction

Recently, the intrinsic magnetic topological insulators have attracted tremendous research interest because they can potentially host a variety of exotic topological quantum states such as the quantum anomalous Hall effect (QAHE) and the axion insulator state.[1, 2] Thanks to the theoretical predictions followed by the successful growth of sizeable single crystals, $MnBi_2Te_4$ (MBT) is now regarded as the first experimental realization of an intrinsic magnetic topological insulator.[3-8] At ambient pressure, MBT crystallizes in the tetradymite-type structure with a rhombohedral space group *R-3m*.[9] The crystal structure consists of Te-Bi-Te-Mn-Te-Bi-Te septuple layers (SLs) that are stacked along the *c*-axis. Each SL is formed by inserting a MnTe bilayer into a $Bi_2Te_3$ quintuple layer. Because these SLs are coupled through van der Waals bonding, atomically thin layers of MBT can be obtained by simple mechanical exfoliation. MBT thus offers a unique natural heterostructure intergrown between magnetic planes and layers of topological insulators.[10]

Transport and magnetic measurements on the MBT single crystals reveal a metallic conductivity with dominant n-type charge carriers and a long-range antiferromagnetic (AF) order at $T_N \approx 24$ K. The temperature dependence of resistivity $\rho(T)$ displays a cusp-like anomaly at $T_N$ followed by a rapid drop upon further cooling.[6, 8, 11] Refinements of powder neutron diffraction data have established an A-type AF structure consisting of ferromagnetic (FM) layers coupled antiferromagnetically with an ordered moment ~4 $\mu_B$/$Mn^{2+}$ pointing along the *c*-axis.[8] Because the interlayer AF coupling is relatively weak, a moderate external magnetic field $H_{c1} \sim 3.3$ T applied along the *c*-axis can induce a spin-flop transition from the A-type AF order to a canted AF (cAF) state. A polarized FM state is eventually achieved at $H_{c2} \sim 7.8$ T with a saturation moment of 3.56 $\mu_B$/Mn. As a result, the Hall resistivity $\rho_{xy}(H)$ of MBT at $T < T_N$ displays a sharp drop at $H_{c1}$ followed a weak anomaly at $H_{c2}$. Thus, these characteristic anomalies in $\rho(T)$ and $\rho_{xy}(H)$ can be used to determine $T_N$ and $H_{c1,c2}$.

During the past few months, several important experimental progresses have been achieved on MBT. For example, Deng *et al.*[12] observed quantized anomalous Hall effect in the thin-flake samples under a moderate magnetic field, while Liu *et al.*[13] realized a quantum phase transition from axion insulator to Chern insulator by applying magnetic field to the exfoliated flake with 6 SLs. In addition, Zheng *et al.*[14] achieved the gate-controlled reversal of anomalous Hall effect in the 5-SLs MBT devices. Moreover, MBT thin films with various thickness can now be grown in a well-controlled manner with molecular beam epitaxy,[15] for which both the archetypical Dirac surface state and intrinsic magnetic order have been observed.

Although the available results on MBT are encouraging, experimental realizations of QAHE remain elusive so far. For an ideal magnetic topological insulator, it is essential to achieve an insulating bulk state, while existing MBT single crystals are metallic with dominated n-type carriers.[8, 16] In addition, single crystals were found to contain considerable amount of Mn/Bi anti-site disorders and/or Mn vacancies, which can alter the transport properties significantly.[6, 11] Depending on the level of defects, either metallic or insulating-like $\rho(T)$ has been observed in the exfoliated thin flakes of MBT.[11] For the thin flakes, gate-voltage tuning has been applied to suppress the bulk carriers and to tune the Fermi level into the surface bang gap so as to observe the above-mentioned quantum phase transitions.[13] On the other hand, substitutions of Sb for Bi in the series of Mn(Sb$_x$Bi$_{1-x}$)$_2$Te$_4$ have also been attempted to open the bulk band gap through shifting the Fermi level.[16] Although a transition from n-to-p type charge carriers has been achieved, metallic conduction remains prevailing in the series of Mn(Sb$_x$Bi$_{1-x}$)$_2$Te$_4$.[16, 17]

During the course of a high-pressure (HP) study, we found that the $\rho(T)$ of MBT single crystal is enhanced progressively with increasing pressure and even changes from metallic to an activated behavior, *i.e.* d$\rho$/d$T$ < 0, at $P$ > ~ 7 GPa. The $\rho(T)$ data at HP even resemble those of gate-voltage-tuned thin flakes with the Fermi level lying inside the surface band gap.[13] Meanwhile, the AF order manifested by the resistivity anomaly exhibits nonmonotonic variation with pressure and seems to be suppressed at about 7-8 GPa. Since no structural transition is observed up to 12.8 GPa, the observed suppression of AF metallic state should be ascribed to a pressure-induced electronic state transition. Based on the Hall resistivity measurements and first-principles calculations, we have discussed some possible origins responsible for these experimental findings.

## Experimental

MBT single crystals used the present study were grown out of a Bi-Te flux and well characterized.[8] The HP resistivity and Hall effect were measured with the standard four-probe method in a palm-type cubic anvil cell (CAC) apparatus. Details about the experimental setup can be found elsewhere.[18] Glycerol was used as the pressure transmitting medium (PTM). The pressure values were estimated from the pressure-load calibration curve determined by observing the characteristic transitions of Bi and Pb at room temperature. The HP structural study was carried out with a diamond-anvil cell (DAC) mounted on a four-circle X-ray diffractometer (Bruker P4) with Mo anode. A small amount of NaCl powder was mixed with the sample to show the pressure inside the chamber filled with a mixture of methanol and ethanol as the PTM. XRD pattern was collected with an image plate from Fujifilm. We used the software FIT2D to integrate the XRD pattern into intensity versus 2θ. The unit-cell parameters were extracted from refining XRD patterns with the LeBail method.

## Results and discussions

### HP resistivity

We performed HP resistivity measurements on two different MBT single crystals,

denoted as #1 and #2 hereafter. The current was applied within the *ab* plane. Figure 1(a) shows the $\rho(T)$ curves of sample #1 under various pressures up to 12.5 GPa. At ambient pressure, $\rho(T)$ exhibits a metallic behavior in the whole temperature range and displays a cusp-like anomaly at the AF transition $T_N$ = 24.7 K marked by an arrow. The rapid drop of resistivity below $T_N$ should be ascribed to the reduction of spin scattering after the formation of long-range AF order.[19] With increasing pressure, interestingly, the magnitude of $\rho(T)$ is enhanced progressively and the temperature dependences display unusual evolutions. For $P$ < 6 GPa, the $\rho(T)$ curves exhibit similar behaviors characterized by a cusp-like anomaly at $T_N$ followed by a drop of resistivity below. As shown by the arrows in Fig. 1(a), $T_N$ determined from the resistivity anomaly first increases to ~29.6 K at 2 GPa, and then gradually decreases to ~18.3 K at 5.5 GPa. Meanwhile, the drop of resistivity below $T_N$ becomes weaker and even changes into a slight upturn at low temperatures at 5.5 GPa. When increasing pressure to 7 GPa, the upturn trend becomes much stronger and no obvious anomaly can be discerned in $\rho(T)$, signaling the possible suppression of the AF order. Eventually, an activated behavior is also realized in the high-temperature region at $P$ > 10 GPa and becomes more evident with increasing pressure.

To further check the variation of $T_N$ with pressure, we performed HP resistivity measurements on the sample #2. As seen in Fig. 1(b), we obtained similar results as sample #1, *i.e.* the $\rho(T)$ curves move up for $P \geq$ 2.5 GPa and the $T_N$ manifested by the resistivity anomaly is suppressed gradually with increasing pressure. In comparison with sample #1, the critical pressure for the absence of resistivity anomaly at $T_N$ in sample #2 is slightly higher, presumably due to some sample-dependent issues. The $\rho(T)$ data at 6 and 7 GPa in Fig. 1(b) illustrate more clearly how the resistivity is enhanced accompanying the suppression of AF order to ~13 K at 7 GPa. For sample #2, we also checked resistivity after releasing pressure. As shown in Fig. 1(b), except for a slight increase of resistivity the sample almost recovers to its initial state with a cusp-like resistivity anomaly at $T_N$. Such a reversible pressure effect should be attributed to the excellent hydrostatic pressure condition in our CAC up to at least 15 GPa.[18] Therefore, our preliminary HP resistivity measurements evidenced the concomitant suppression of AF order and development of an activated resistivity behavior in the pressurized MBT single crystal.

### *T-P* phase diagram

Based on the above resistivity measurements, we tentatively construct a temperature-pressure phase diagram for MBT single crystal as shown in Fig. 2. A contour plot of d$\rho$/d$T$ is also superimposed to highlight the evolutions of the dominant conduction mechanism. As can be seen, the AF transition temperature $T_N$ is at first slightly enhanced by pressure of ~ 2 GPa, and then suppressed gradually until vanished completely at ~7 GPa. Accompanying the suppression of AF order, the electrical transport properties at low temperatures also change dramatically from metallic

d$\rho$/dT > 0 to an activated, semiconducting-like behavior of d$\rho$/dT < 0, as illustrated by the color coding in Fig. 2. In the paramagnetic state above $T_N$, a mild crossover from metallic to semiconducting behavior also takes place at higher pressures. Although the pressure-induced suppression of AF order is not unexpected, the observed monotonic enhancement of resistivity and the activated behavior at high pressures are counterintuitive in the sense that pressure usually broadens the electronic bandwidth.

**HP Hall resistivity**

In order to gain more insights into the evolution of electronic state under pressure, we further measured Hall resistivity $\rho_{xy}(H)$ on the MBT sample #2, which was recovered from the above HP resistivity measurements. In addition to the carrier information, $\rho_{xy}(H)$ data may also provide some hints about the magnetic state through detecting the anomalies caused by the spin-flop transitions mentioned above. We have measured $\rho_{xy}(H)$ in between +8.5 T and -8.5 T at fixed temperatures of 1.5, 10, 20, and 30 K under pressures of 1, 3, 5 and 8 GPa. The current was applied within the *ab* plane and the magnetic field applied along the *c*-axis. The $\rho_{xy}(H)$ data collected for a field sweeping between +8.5 and -8.5 T are anti-symmetrized in order to eliminate the contributions from the longitudinal magnetoresistance.

Some representative $\rho_{xy}(H)$ data at 1.5 K (< $T_N$) and 30 K (> $T_N$) under different pressures are shown in Figs. 3(a, b). All $\rho_{xy}(H)$ curves display an initial negative slope, suggesting that the electron-type carriers dominate the transport properties up to at least 8 GPa. For $T$ = 1.5 K, Fig. 3(a), $\rho_{xy}(H)$ at 1 GPa exhibits a gradual drop starting at 4.8 T and ending at 5.3 T due to the field-induced spin-flop transition from AF to cAF. In comparison with the sharp drop of $\rho_{xy}(H)$ at $H_{c1}$ ~ 3.3 T at ambient pressure,[8] the spin-flop transition seen here at 1 GPa takes place not only at a larger magnetic field but also in a rather retarded fashion. If the single-ion anisotropy does not vary substantially, these observations indicate that the interlayer AF coupling is strengthened due to the reduction of *c*-axis. This is also consistent with the initial enhancement of $T_N$ seen in Fig. 2. As a result, we cannot detect the cAF to FM transition in $\rho_{xy}(H)$ up to 8.5 T, the highest field in our measurements.

When increasing pressure to ≥ 3 GPa, the initial slope of $\rho_{xy}(H)$ decreases and keeps nearly constant, implying an enhanced carrier density as discussed below. For $P$ = 3 GPa, the spin-flop transition occurs at much higher field of $H_{c1}$ ≥ 7 T, and becomes much more broaden than that at 1 GPa. In addition, the magnitude of $\rho_{xy}(H)$ drop at the transition is also reduced considerably. The continuous increase of $H_{c1}$ at 3 GPa would suggest a further enhancement of interlayer AF coupling and thus $T_N$. However, the observed reduction of $T_N$ at 3 GPa in Figs. 1 and 2 implies that some other competing factors are at play under HP. Under the hydrostatic pressure conditions, both *a* and c axes are compressed, and thus both intralayer and interlayer magnetic interactions are expected to be modified accordingly. It has been proposed in the MnBi$_{2-x}$Sb$_x$Te$_4$ that the reduced nearest-neighbor Mn-Mn distance within the *ab* plane upon Sb doping would increase the direct AF interactions that can compete with the dominant intralayer FM interactions.[17] In the present case, HP may also enhance the intralayer direct AF interactions and thus magnetic competition, leading to the gradual reduction of $T_N$ above 2 GPa in Fig. 2. Meanwhile, such a magnetic competition should also reduce the tendency for FM alignment of spins, which can rationalize the reduced magnitude of $\rho_{xy}(H)$ drop at the spin-flop transition of 3 GPa. For $P$ = 5 and 8 GPa, $\rho_{xy}(H)$ curves are perfect linear without any anomaly up to 8.5 T. Further studies in a larger field range are needed to address whether the spin-flop transition occurs at $H_{c1}$ much higher than 8.5 T or disappears gradually as the intralayer AF interaction becomes dominant. For $T$ = 30 K, Fig. 3(b), all $\rho_{xy}(H)$ curves are linear in field up to 8.5 T and the negative slope is reduced gradually with increasing pressure.

We have extracted the Hall coefficient $R_H$ from a linear fitting to the $\rho_{xy}(H)$ curves in the linear region as indicated by the dotted lines in Figs. 3(a, b). The obtained $R_H(P)$ at 1.5 and 30 K are plotted in Fig. 3(c). For both temperatures, $|R_H(P)|$ decreases quickly and then tends to level off above 3 GPa. This observation indicates that the influence of pressure on the electronic state is moderate for $P$ > 3 GPa. Based on a single-band model, we estimated the carrier density $n = -(eR_H)^{-1}$ and mobility $\mu = R_H/\rho$, which are displayed in Figs. 3(d, e). At 1.5 K, the obtained carrier density at 1 GPa ~$1.2\times10^{20}$ cm$^{-3}$ is slightly large than that of $0.9 \times10^{20}$ cm$^{-3}$ at ambient pressure,[8] and is further increased to ~$1.8\times10^{20}$ cm$^{-3}$ above 3 GPa. The carrier density at 30 K also increases progressively from $0.94\times10^{20}$ cm$^{-3}$ at 1 GPa to $2.16\times10^{20}$ cm$^{-3}$ at 8 GPa. Since the carrier density is improved by pressure, the observed continuous enhancement of resistivity and the activated behaviors at low temperature and higher pressures cannot be attributed to suppression of bulk carriers or the opening of bulk band gap. Instead, it should be ascribed to the significant reduction of carrier mobility shown in Fig. 3(e). As discussed below, the carrier localization might correlated with the suppression of AF order due to the enhanced magnetic competitions.

It is also noteworthy that $R_H(P)$ at 1.5 and 30 K are crossed at 5 GPa as illustrated by opposite directions of the vertical arrows in Fig. 3(c). This fact reflects the distinct electronic states of MBT at low and high pressures. As shown in Fig. 3(g), $n(T)$ at 1 GPa experience a sudden enhancement below ~ 20 K, corresponding to the rapid drop of resistivity below $T_N$ in Fig. 1. Similar feature is still visible at 3 and 5 GPa, but it becomes much weaker, which is also consistent with the diminishing resistivity drop below $T_N$. In contrast, $n(T)$ at 8 GPa monotonically decrease upon cooling, which also explains the observed resistivity upturn or activated behavior at low temperatures under $P$ > 7 GPa. Therefore, these systematic investigations on the HP Hall resistivity further elaborate a pressure-induced electronic transition in MBT single crystal.

### HP structural study

Before discussing the possible origin for this pressure-induced transition, we first investigated the structural responses under pressure. No obvious sudden changes were observed in the above HP resistivity measurements during the compression process, implying the absence of pressure-induced structural transition at least up to 12.5 GPa. In order to confirm this, we performed HP XRD measurements at room temperature. Figure 4(a) shows the XRD patterns of MBT up to 12.8 GPa. The observation of similar patterns confirms the absence of structural phase transition in the investigated pressure range. It should be noted that some peaks become rather weak under pressure due to the development of preferred orientation for this layered material. The lattice parameters, *a*, *c*, and *V*, were extracted from these XRD patterns with the LeBail method, and are displayed in Figs. 4(b, c) as a function of pressure. As can be seen, all unit-cell parameters decrease monotonically with pressure and no clear anomaly can be discerned in the studied pressure range. It was also found that the unit cell experiences a strong anisotropic compression, i.e. $\Delta a/a_0$ = -4.77 % versus $\Delta c/c_0$ = -8.91 %, up to 12.8 GPa. This is mainly due to the layered structure of MBT with weak van der Waals bonding along the *c*-axis. The smooth evolution of volume with pressure can be well described with the Birch-Murnaghan equation, *i.e.*

$$P = \frac{3B_0}{2}\left[\left(\frac{V_0}{V}\right)^{7/3} - \left(\frac{V_0}{V}\right)^{5/3}\right]\left\{1 + \frac{3}{4}(B'-4)\left[\left(\frac{V_0}{V}\right)^{2/3} - 1\right]\right\} \quad (1),$$

The best fit yields a bulk modulus $B_0$ = 25.6 ± 4.8 GPa, $B'$ = 11.7 ± 2.2, and $V_0$ = 662.6 ± 4.8 Å³, respectively. The obtained $B_0$ of MnBi$_2$Te$_4$ is close but smaller than that of 35(2) GPa for SnBi$_2$Te$_4$.[20]

Since our HP XRD measurements ruled out the structural transition in the studied pressure range, the observed suppression of AF metallic state should be caused by some electronic state change induced by lattice compression. As mentioned above, the carrier density is actually raised by pressure, Fig. 3(d), thus the enhancement of resistivity under pressure cannot be explained by the suppression of bulk carrier or band gap opening. Alternatively, the charge carriers should experience a strong tendency for localization, which is correlated closely with the evolution of magnetism. This is supported by the fact that both the development of resistivity upturn at low temperature and the activated behavior in the high temperature region at high pressures are accompanied with the suppression of long-range AF order as seen in Fig. 1. As pointed out above, reduction of unit-cell parameters under HP not only strengthens the interlayer AF coupling that leads to the initial increase of $T_N$ but also enhances the intralayer direct AF interactions, which can compete with the dominant FM interactions. Such an AF/FM competition is expected to give rise to a strong magnetic frustration that would prevent the formation of long-rang AF order. This can explain the decrease and final suppression of $T_N$ above 2 GPa in Fig. 2. In addition, the magnetic frustration in an itinerant-electron system should also enhance the electron scattering and promote charge localizations. This effect becomes much stronger at $P$ > 7 GPa when the long-range AF order cannot form and magnetic fluctuations becomes prevailing due to the presence of strong magnetic frustrations.

As such we end up with an activated behavior in resistivity, presumably through magnetic fluctuations induced carrier localization process, even though the magnitude of resistivity remains in the bad metal regime. Similar behavior has been observed in the geometrically frustrated AF pyrochlore $Ca_2Ru_2O_7$ due to the interplay between frustrated magnetism and the itinerant electron.[21] A HP neutron study is needed to prove the suppression of AF order.

**First-principles calculations**

To shed more light on the above experimental finding, we have performed first-principles calculations using the augmented plane-wave all-electron code WIEN2K [22] within the generalized gradient approximation of Perdew, Burke and Ernzerhof.[23] These calculations were conducted at the unit-cell parameters corresponding to ambient pressure, 7 GPa, and an effective pressure of ~20 GPa with $a$ = 4 Å and 36 Å, respectively. Because it is well known that the suppression of magnetic order by pressure is often underestimated by first-principles approaches, we have chosen a rather large pressure as a sensitive check on the calculations. At ambient pressure, we find the A-type AF state to fall ~7 meV per Mn below the FM configuration, in agreement with the experimental result. Note that our ambient-pressure calculations without spin-orbit coupling find a band gap of ~ 0.63 eV and a calculated Mn staggered moment of ~4.29 $\mu_B$ for this A-type AF state. On the other hand, a metallic behavior is obtained in a non-spin-polarized calculation.

With the application of 7 GPa pressure, the calculated staggered moment is slightly reduced to 4.12 $\mu_B$, and the AF state shows a semi-metallic behavior with a band overlap of about 50 meV. At the assumed 20 GPa pressure, however, we find a much more substantial reduction in magnetic order, with a calculated moment of 3.59$\mu_B$. Note a small fraction of this reduction is due to the use of smaller "muffin-tin" radii, within which the Mn spin moment is calculated, in this highly compressed unit cell. This larger reduction is at least qualitatively consistent with the observed suppression of magnetic order by pressure. Note that we have not assessed the effects of pressure on the exchange interactions ultimately determining the ordering point. Similar with the 7 GPa case, we find a metallic behavior at 20 GPa with a much larger band overlap of e ~ 300 meV. The increased band overlap seen from calculations is consistent with the observed enhancement of carrier density seen in Fig. 3(d).

The theoretical finding of metallic behavior under applied pressure, but semiconducting behavior at ambient pressure, seems to contradict the experimental transport results that show the reverse. However, there are numerous complications affecting the comparison between theory and experiment. First and foremost, the disappearance of long-range magnetic order at $T_N$ does not necessarily signal the loss of all magnetic character. One plausible cause is the magnetic frustration due to the AF/FM competition as discussed above. In addition, for a highly anisotropic material such as MBT with large local moment, it is also likely that local magnetic correlations persist well above the nominal ordering point.[24] Such behavior is evident in numerous other 3*d*-based magnetic materials, such as $Mn_3Si_2Te_6$ [25] and $LiGaCr_4S_8$ [26], where magnetism-related semiconducting behavior persists to temperatures well above the long-range order temperature. Although the complex behavior of the observed resistivity with pressure most likely originates in all these factors, further experimental and theoretical study will be needed to develop a detailed understanding.

## Conclusion

In summary, we have performed a comprehensive high-pressure study on the MBT single crystal, which is considered as the first intrinsic antiferromagnetic topological insulator. We found that its resistivity is gradually enhanced by pressure and even changes from metallic to activated behavior at *P* >7 GPa. In addition, we found that the AF order was initially strengthened due to the reduction of interlayer distances, but then suppressed gradually until vanished completely at ~ 7-8 GPa. The layered structure of MBT is confirmed to preserve at pressures up to at least 12.8 GPa. We have discussed our experimental findings in terms of the magnetic competition based on the results of Hall resistivity measurements and first-principles calculations. Our results call for further experimental and theoretical studies in order to achieve a better understanding on the interplay between magnetism and transport properties of MBT.

## Acknowledgements

This work is supported by the National Key R&D Program of China (Grant Nos. 2018YFA0305700 and 2018YFA0305800), the National Natural Science Foundation of China (Grant Nos. 11888101, 11574377, 11834016, 11874400), the Strategic Priority Research Program and Key Research Program of Frontier Sciences of the Chinese Academy of Sciences (Grant Nos. XDB25000000, XDB07020100 and Grant No. QYZDB-SSW-SLH013). JQY and DSP are supported at the Oak Ridge National Laboratory by the US Department of Energy, Office of Science, Basic Energy Sciences, Materials Sciences and Engineering Division. J.S.Z. is supported by the National Science Foundation DMR-1729588.

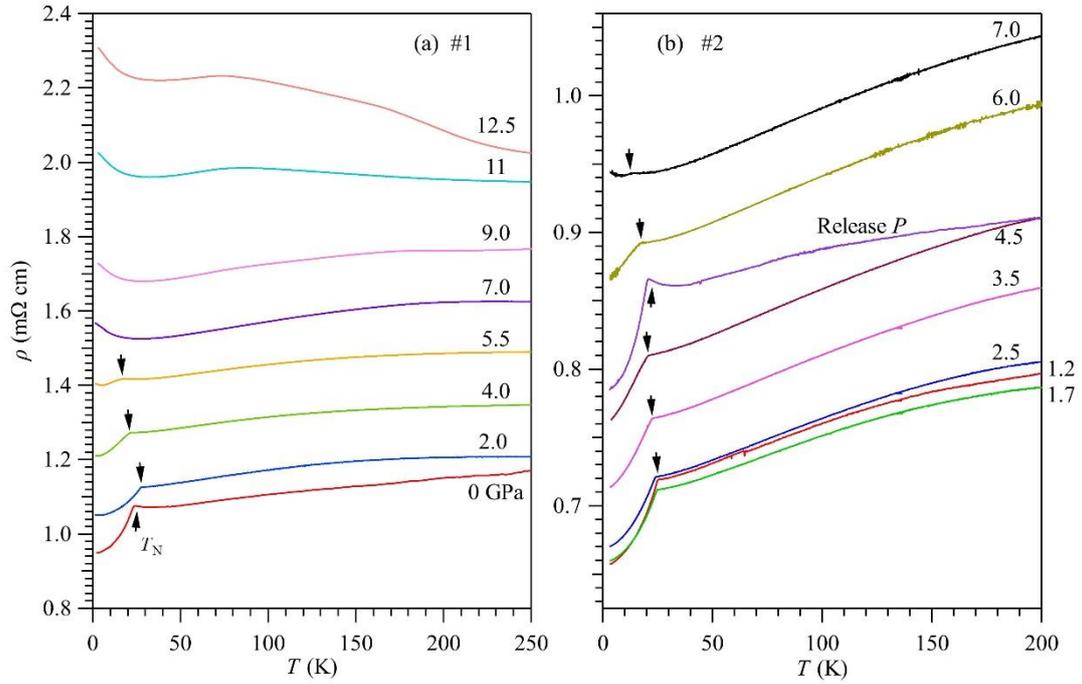

Figure 1. Temperature dependence of resistivity $\rho(T)$ for MnBi$_2$Te$_4$ under various pressures up to 12.5 GPa for sample #1(a) and up to 7 GPa for sample #2(b). The Neel temperature $T_N$ is marked by the black arrow.

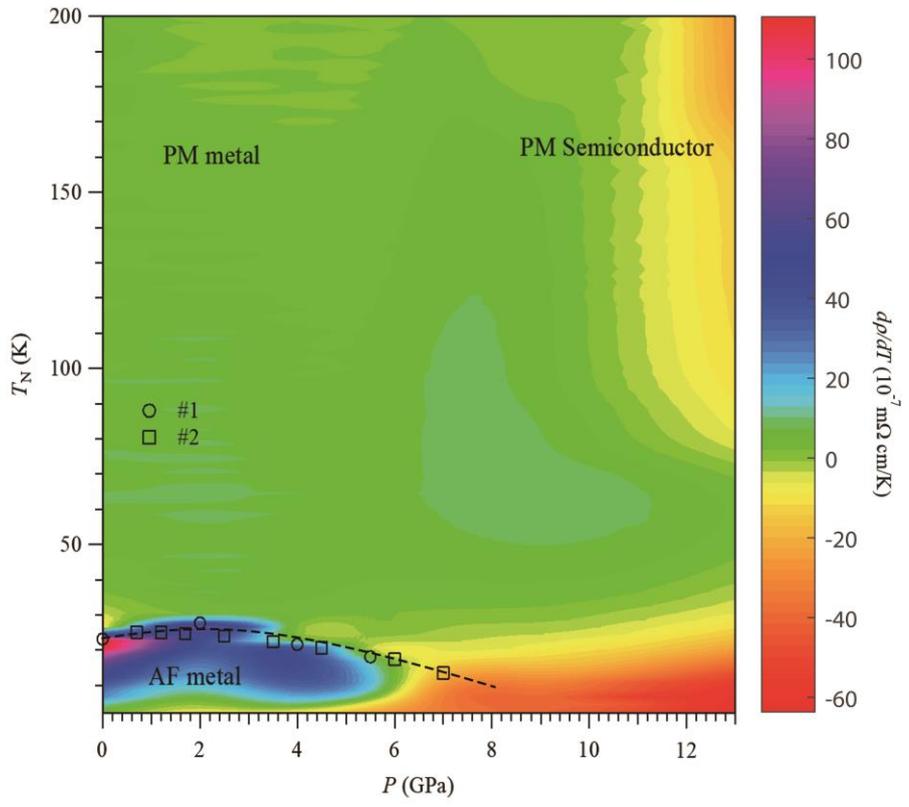

Figure 2. Temperature-pressure phase diagram of MnBi$_2$Te$_4$.

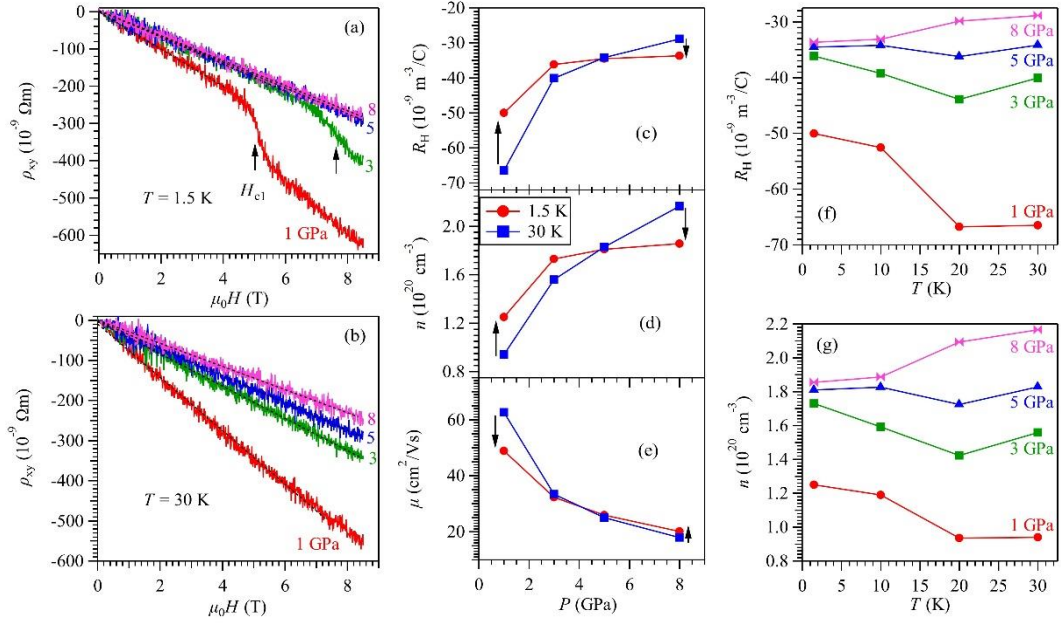

Figure 3. (a, b) Field dependence of Hall resistivity $\rho_{xy}(H)$ of MBT #2 at 1.5 K and 30 K under different pressures. (c, d, e) Pressure dependences of Hall coefficient $R_H$, carrier density $n$, and mobility $\mu$ at 1.5 and 30 K. (f, g) Temperature dependences of $R_H$ and $n$ at different pressures.

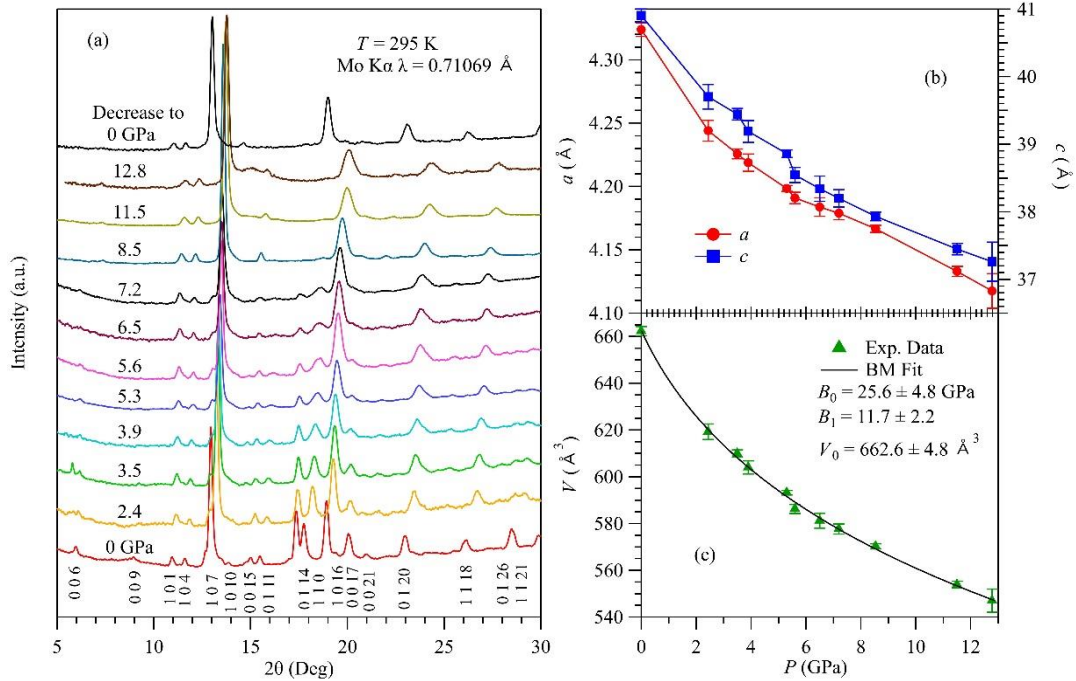

Figure 4. (a) Powder XRD patterns of $MnBi_2Te_4$ up to 12.8 GPa, (b, c) pressure dependence of the unit-cell parameters $a$, $c$, and $V$.